\documentstyle[12pt]{article}
\author{\small H.T. Nieh and Gang Su\thanks{On leave of absence from the
Graduate School, Chinese Academy of Sciences, Beijing, China.}\\
\small Institute for Theoretical Physics\\
\small State University of New York at Stony Brook\\
\small Stony Brook, New York 11794, USA
\and
\small Bao-Heng Zhao\\
\small Department of Physics, Graduate School\\
\small Chinese Academy of Sciences, P. O. Box 3908\\
\small Beijing 100039, China}
\title{\large \bf Off-Diagonal Long-Range Order: Meissner Effect and Flux
Quantization
}
\date{}
\begin{document}
\maketitle
\begin{abstract}
There has been a proof by Sewell that the hypothesis of off-diagonal
long-range order in the reduced density matrix $\rho _2$ implies the
Meissner effect. We present in this note an elementary and straightforward
proof that not only the Meissner effect but also the property of magnetic
flux quantization follows from the hypothesis. It is explicitly shown that
the two phenomena are closely related, and phase coherence is the origin for
both.

\end{abstract}

\newpage
\centerline{\bf Introduction}

Penrose and Onsager$^1$ first proposed that the mathematical statement of
Bose-Einstein condensation in the case of interacting boson systems is the
existence of off-diagonal long-range order (ODLRO) in the reduced density
matrix $\rho _1$. Yang$^2$ later made a thorough study of the subject and
generalized the concept to interacting fermion systems. It is now generally
accepted that the superconducting state of electrons is characterized by the
existence of off-diagonal long-range order in the reduced density matrix $%
\rho _2$.

In view of the fundamental importance of the concept of ODLRO, it is
important that there be a rigorous proof of the assertion that ODLRO in the
reduced density matrix $\rho _2$ implies both the Meissner effect and the
quantization of magnetic flux, which are the basic characteristic properties
of superconducting states. There has been a recent proof by Sewell$^3$ of
the Meissner effect as a consequence of ODLRO, under certain simplifying
assumptions. We shall extend in this note the proof to the case of magnetic
flux quantization.

The proof we present is simple and elementary, and clearly shows that both
the Meissner effect and the flux quantization are closely related phenomena,
and the underlying origin for both is the coherence of phases. \\

\centerline{\bf System in Uniform Magnetic Field}

We consider a system of $N$ electrons inside a crystal, which is placed in a
uniform external magnetic field. We make the simplifying assumption that the
magnetic field ${\bf B}$ inside the material is also uniform, with the
corresponding vector potential, given by ${\bf A}({\bf r})$, which is of the
general form:
$$
{\bf A}({\bf r})={\bf A}_0({\bf r})+\nabla \phi ({\bf r}),\eqno(1)
$$
where ${\bf A}_0({\bf r})=\frac 12{\bf B}\times {\bf r,}$ and the gauge $%
\phi ({\bf r})$ is, in general, multi-valued in ${\bf r}$. The Hamiltonian
for the system is
$$
H=\sum_j\frac 1{2m}[\frac \hbar i\nabla _j+\frac ec{\bf A}({\bf r}%
_j)]^2+\frac 12\sum_{j,k}V(r_{jk})+\sum_jV_c({\bf r}_j),\eqno(2)
$$
where $V_c({\bf r}_j)$ represents the interaction of the crystal lattice on
the j-th electron, and the two-particle potential $V(r_{jk})$, which
includes the effective interaction due to phonon exchange, depends on the
distance between the two electrons. Let
$$
\psi _n({\bf r}_1,{\bf r}_2,\cdots ,{\bf r}_N),~~~~~n=1,2,...,
$$
represent a complete orthonormal set of single-valued and properly
antisymmetrized eigenfunctions of the Hamiltonian
$$
H\psi _n({\bf r}_1,{\bf r}_2,\cdots ,{\bf r}_N)=E_n\psi _n({\bf r}_1,{\bf r}%
_2,\cdots ,{\bf r}_N).\eqno(3)
$$
We neglect the edge effects near the boundaries, and assume the usual
periodic boundary conditions for the wave functions. The phenomenon of flux
penetration near boundaries is beyond the scope of the present idealized
considerations. \\

\centerline{\bf Two Complete Sets of Eigenfunctions}

Now, consider a space displacement
$$
{\bf r}_j\rightarrow {\bf r}_j-{\bf a},
$$
under which
$$
V(r_{jk})\rightarrow V(r_{jk}),
$$
$$
{\bf A}({\bf r}_j)\rightarrow {\bf A}({\bf r}_j-{\bf a})={\bf A}({\bf r}_j)+%
{\bf \nabla }_j[{\bf a}\cdot {\bf A}_0({\bf r}_j)+\phi ({\bf r}_j-{\bf a}%
)-\phi ({\bf r}_j)].\eqno(4)
$$
We observe that a space displacement induces a gauge transformation in the
vector potential. It is this property that makes it possible for a simple
proof of the flux quantization and the Meissner effect as consequences of
the ODLRO hypothesis. The lattice potential $V_c({\bf r}_j)$ has the
symmetry of the lattice, and is invariant under a displacement only if the
displacement respects the symmetry of the crystal lattice.

We first consider the case where the lattice potential $V_c({\bf r}_j)$ in
the Hamiltonian is omitted, and will come back later to consider the case
where it is not. Without the lattice potential, the eigen-equation (3),
under the displacement ${\bf r}_j\rightarrow {\bf r}_j-{\bf a}$, becomes
$$
\{\sum_j\frac 1{2m}[\frac \hbar i\nabla _j+\frac ec{\bf A}({\bf r}_j-{\bf a}%
)]^2+\frac 12\sum_{j,k}V(r_{jk})\}\psi _n({\bf r}_1-{\bf a},{\bf r}_2-{\bf a}%
,\cdots ,{\bf r}_N-{\bf a})
$$
$$
=E_n\psi _n({\bf r}_1-{\bf a},{\bf r}_2-{\bf a},\cdots ,{\bf r}_N-{\bf a}).%
\eqno(5)
$$
Noting that a displacement induces a gauge transformation, as in (4), in the
vector potential, we can absorb the gauge into a phase for the electron wave
function to obtain
$$
\{\sum_j\frac 1{2m}[\frac \hbar i\nabla _j+\frac ec{\bf A}({\bf r}%
_j)]^2+\frac 12\sum_{j,k}V(r_{jk})\}\exp [\frac{ie}{c\hbar }\sum_j\chi _{%
{\bf a}}({\bf r}_j)]\psi _n({\bf r}_1-{\bf a},{\bf r}_2-{\bf a},\cdots ,{\bf %
r}_N-{\bf a})
$$
$$
=E_n\exp [\frac{ie}{c\hbar }\sum_j\chi _{{\bf a}}({\bf r}_j)]\psi _n({\bf r}%
_1-{\bf a},{\bf r}_2-{\bf a},\cdots ,{\bf r}_N-{\bf a}),\eqno(6)
$$
where
$$
\chi _{{\bf a}}({\bf r}_j)={\bf a}\cdot {\bf A}_0({\bf r}_{j})+\phi ({\bf r}%
_j-{\bf a})-\phi ({\bf r}_j).\eqno(7)
$$
It is seen that $\psi _n^{\prime }({\bf r}_1,{\bf r}_2,\cdots ,{\bf r}_N)$,
given by
$$
\psi _n^{\prime }({\bf r}_1,{\bf r}_2,\cdots ,{\bf r}_N)=\exp [\frac{ie}{%
c\hbar }\sum_j\chi _{{\bf a}}({\bf r}_j)]\psi _n({\bf r}_1-{\bf a},{\bf r}_2-%
{\bf a},\cdots ,{\bf r}_N-{\bf a}),\eqno(8)
$$
are again eigenfunctions of the original Hamiltonian. In the following we
will only consider the case of {\bf a} to be infinitesimal on macroscopic
scale$^4.$ In this case the phase $\chi _{{\bf a}}({\bf r})$ is
single-valued in ${\bf r.}$ Therefore, $\psi _n^{\prime }$ form another
complete orthonormal set of single-valued eigenfunctions. We thus have two
bases of eigenfunctions, $\psi _n$ and $\psi _n^{\prime }$. \\

\centerline{\bf Reduced Density Matrix for Two Particles}

The reduced density matrix element $\rho _2({\bf r^{\prime }}_1,{\bf %
r^{\prime }}_2;{\bf r}_1,{\bf r}_2)$ is defined in terms of a complete
orthonormal set of energy eigenfunctions by
$$
\rho _2({\bf r^{\prime }}_1,{\bf r^{\prime }}_2;{\bf r}_1,{\bf r}_2)=\int
\cdots \int \frac{d{\bf r}_3\cdots d{\bf r}_N}{(N-2)!}\frac 1Z\sum_n\exp
(-E_n/kT)
$$
$$
\times \psi _n({\bf r^{\prime }}_1,{\bf r^{\prime }}_2,{\bf r}_3,\cdots ,%
{\bf r}_N)\psi _n^{*}({\bf r}_1,{\bf r}_2,{\bf r}_3,\cdots ,{\bf r}_N),%
\eqno(9)
$$
where $Z$ is the partition function, and the normalization is such that
$$
Tr\rho _2=\int \int \rho _2({\bf r}_1,{\bf r}_2;{\bf r}_1,{\bf r}_2)d{\bf r}%
_1d{\bf r}_2=N(N-1).
$$
We note that $\rho _2({\bf r^{\prime }}_1,{\bf r^{\prime }}_2;{\bf r}_1,{\bf %
r}_2)$ is single-valued, as the wave-functions are single-valued. Since the
reduced density matrices are basis independent, we can equally represent \\ $%
\rho _2({\bf r^{\prime }}_1,{\bf r^{\prime }}_2;{\bf r}_1,{\bf r}_2)$ in
terms of the primed set of energy eigenfunctions:
$$
\rho _2({\bf r^{\prime }}_1,{\bf r^{\prime }}_2;{\bf r}_1,{\bf r}_2)=\int
\cdots \int \frac{d{\bf r}_3\cdots d{\bf r}_N}{(N-2)!}\frac 1Z\sum_n\exp
(-E_n/kT)
$$
$$
\times \psi _n^{\prime }({\bf r^{\prime }}_1,{\bf r^{\prime }}_2,{\bf r}%
_3,\cdots ,{\bf r}_N)\psi _n^{\prime }{}^{*}({\bf r}_1,{\bf r}_2,{\bf r}%
_3,\cdots ,{\bf r}_N),\eqno(10)
$$
Expressing the primed eigenfunctions in the above equation in terms of the
original eigenfunctions, we obtain
$$
\rho _2({\bf r^{\prime }}_1,{\bf r^{\prime }}_2;{\bf r}_1,{\bf r}_2)=\exp \{
\frac{ie}{c\hbar }[\chi _{{\bf a}}({\bf r^{\prime }}_1)+\chi _{{\bf a}}({\bf %
r^{\prime }}_2)-\chi _{{\bf a}}({\bf r}_1)-\chi _{{\bf a}}({\bf r}_2)]\}
$$
$$
\times \int \cdots \int \frac{d{\bf r}_3\cdots d{\bf r}_N}{(N-2)!}\frac
1Z\sum_n\exp (-E_n/kT)
$$
$$
\times \psi _n({\bf r^{\prime }}_1-{\bf a},{\bf r^{\prime }}_2-{\bf a},{\bf r%
}_3-{\bf a},\cdots ,{\bf r}_N-{\bf a})\psi _n^{*}({\bf r}_1-{\bf a},{\bf r}%
_2-{\bf a},{\bf r}_3-{\bf a},\cdots ,{\bf r}_N-{\bf a}).
$$
On account of the periodic boundary conditions satisfied by the wave
functions, we can make an innocent shift of the integration variables in the
integral in the above equation, resulting in the relation of the density
matrix at different space points:
$$
\rho _2({\bf r^{\prime }}_1,{\bf r^{\prime }}_2;{\bf r}_1,{\bf r}_2)=\exp \{
\frac{ie}{c\hbar }[\chi _{{\bf a}}({\bf r^{\prime }}_1)+\chi _{{\bf a}}({\bf %
r^{\prime }}_2)-\chi _{{\bf a}}({\bf r}_1)-\chi _{{\bf a}}({\bf r}_2)]\}\rho
_2({\bf r^{\prime }}_1-{\bf a},{\bf r^{\prime }}_2-{\bf a};{\bf r}_1-{\bf a},%
{\bf r}_2-{\bf a}).\eqno(11)
$$
This equation, as we shall see, is of basic importance. \\

\centerline{\bf Off-Diagonal Long-Range Order}

We can represent the reduced density matrix element in the spectral form:
$$
\rho _2({\bf r^{\prime }}_1,{\bf r^{\prime }}_2;{\bf r}_1,{\bf r}%
_2)=\sum_s\alpha _s\Phi _s({\bf r^{\prime }}_1,{\bf r}_2^{\prime })\Phi
_s^{*}({\bf r}_1,{\bf r}_2),\eqno(12)
$$
where $\alpha _s$ are eigenvalues, and $\Phi _s({\bf r}_1,{\bf r}_2)$ the
properly normalized eigenfunctions of $\rho_2$:
$$
\int \int \Phi _s^{*}({\bf r}_1,{\bf r}_2)\Phi _s({\bf r}_1,{\bf r}_2)d{\bf r%
}_1d{\bf r}_2=1.
$$
We are interested in the property of $\rho _2$ in the off-diagonal
long-range (ODLR) limit:
$$
|{\bf r^{\prime }}_j-{\bf r}_k|\rightarrow \infty ,~~~~~j,k=1,2
$$
while keeping $|{\bf r}_1-{\bf r}_2|$ and $|{\bf r^{\prime }}_1-{\bf %
r^{\prime }}_2|$ finite. The hypothesis of ODLRO in $\rho _2$ is the
statement that in the ODLR limit the reduced density matrix element
factorizes in the form
$$
\rho _2({\bf r^{\prime }}_1,{\bf r^{\prime }}_2;{\bf r}_1,{\bf r}%
_2)\rightarrow \alpha _0\Phi ({\bf r^{\prime }}_1,{\bf r^{\prime }}_2)\Phi
^{*}({\bf r}_1,{\bf r}_2),\eqno(13)
$$
where $\alpha _0$ is the largest eigenvalue of $O(N)$, and $\Phi $ the
corresponding eigenfunction$^2$.

Applying the factorization property of the ODLRO hypothesis to both sides of
the relation (11), we obtain, in the ODLR limit,
$$
\Phi ({\bf r^{\prime }}_1,{\bf r^{\prime }}_2)\Phi ^{*}({\bf r}_1,{\bf r}%
_2)=
$$
$$
\exp \{\frac{ie}{c\hbar }[\chi _{{\bf a}}({\bf r^{\prime }}_1)+\chi _{{\bf a}%
}({\bf r^{\prime }}_2)-\chi _{{\bf a}}({\bf r}_1)-\chi _{{\bf a}}({\bf r}%
_2)]\}\Phi ({\bf r^{\prime }}_1-{\bf a},{\bf r^{\prime }}_2-{\bf a})\Phi
^{*}({\bf r}_1-{\bf a},{\bf r}_2-{\bf a}),\eqno(14)
$$
which implies$^5$
$$
\Phi ({\bf r}_1,{\bf r}_2)=f({\bf a})\exp \{\frac{ie}{c\hbar }[\chi _{{\bf a}%
}({\bf r}_1)+\chi _{{\bf a}}({\bf r}_2)]\}\Phi ({\bf r}_1-{\bf a},{\bf r}_2-%
{\bf a}),\eqno(15)
$$
where $f({\bf a})$ is a possible
displacement dependent, but position independent, phase factor$^6$, with $%
f(0)=1$. Thus, the function $\Phi $, which is to be identified with the
Ginsburg-Landau wave function, at different space points is related by a
phase factor. Repeating successively for infinitesimal displacements along a
closed path C, and picking up products of phase factors and the
path-dependent f-factors, we obtain from (15) the relation
$$
\Phi ({\bf r}_1,{\bf r}_2)=F(C)\exp \{i\frac{2e}{c\hbar }\oint_C{\bf A}({\bf %
r})\cdot d{\bf r}\}\Phi ({\bf r}_1,{\bf r}_2),\eqno(16)
$$
where $F(C)$ represents the product of the f-factors, and use has been made
of the single-valuedness of the function $\Phi ({\bf r}_1,{\bf r}_2)$, a
consequence of the single-valuedness of the density matrix. \\

\centerline{\bf Meissner Effect}

We now use (15) to prove the Meissner effect. In a simply connected region
we apply (15) to two sequences of successive displacements$^7$, first ${\bf a%
}$ followed by ${\bf b}$, and then ${\bf b}$ followed by ${\bf a}$.
Comparing the two resulting relations, we obtain
$$
\exp \{\frac{ie}{c\hbar }[\chi _{{\bf a}}({\bf r}_1)+\chi _{{\bf b}}({\bf r}%
_1-{\bf a})+\chi _{{\bf a}}({\bf r}_2)+\chi _{{\bf b}}({\bf r}_2-{\bf a}%
)]\}=
$$
$$
\exp \{\frac{ie}{c\hbar }[\chi _{{\bf b}}({\bf r}_1)+\chi _{{\bf a}}({\bf r}%
_1-{\bf b})+\chi _{{\bf b}}({\bf r}_2)+\chi _{{\bf a}}({\bf r}_2-{\bf a})]\}.%
\eqno(17)
$$
Namely,
$$
\exp \{i\frac{2e}{c\hbar }{\bf \oint_{(a\times b)}A}({\bf r})\cdot d{\bf r}%
\}=1,\eqno(18)
$$
where the loop path of integration is a parallelogram formed by ${\bf a}$
and ${\bf b}$. Therefore,
$$
{\bf B}\cdot ({\bf a}\times {\bf b)}=n\frac{ch}{2e},~~~n=integers.\eqno(19)
$$
While ${\bf a}$ and ${\bf b}$ can be continuously changed, the right-hand
side of (19), being proportional to an integer, can not. Consistency then
requires
$$
{\bf B}=0.\eqno(20)
$$
This is the property of the Meissner effect. \\

\centerline{\bf Flux Quantization}

Next, we use (15) and (16) to prove the quantization of flux, when the loop $%
C$ is not simply-connected, like a path winding around a tunnel region
inaccessible to the electrons.

As we have already proved the Meissner effect, namely ${\bf B}=0$, it is
clear from (16) that $F(C)=1$ if $C$ is simply-connected. To prove the
quantization of flux, we need to prove $F(C)=1$ for multiply-connected loops
$C$.

Now, the factor $f({\bf a})$ in (15) is independent of ${\bf r}$. This
implies that the property of $F(C)$ is independent of the location of the
loop $C$. Corresponding to a multiply-connected loop $C$, we can consider a
loop of the same shape in a simply-connected region, for which $F(C)=1$. It
follows that $F(C)=1$ also for $C$ in a multiply-connected region. In
general, therefore,
$$
F(C)=1\eqno(21)
$$
for any loop path $C$. (16) then becomes
$$
\Phi ({\bf r}_1,{\bf r}_2)=\exp \{i\frac{2e}{c\hbar }\oint_C{\bf A}({\bf r}%
)\cdot d{\bf r}\}\Phi ({\bf r}_1,{\bf r}_2).\eqno(22)
$$
Consequently, there is the quantization condition
$$
\oint_C{\bf A}({\bf r})\cdot d{\bf r}=n\frac{ch}{2e},~~~~n=integers,\eqno(23)
$$

The loop path $C$ may or may not be simply-connected. When $C$ in (23) is
simply-connected, it can be altered infinitesimally and continuously,
implying, again, the Meissner effect. When $C$ is not simply-connected, like
a path winding around an inaccessible tunnel, the path can not
be deformed across this inaccessible region. Then the quantization condition
(23) implies that the total flux$^8$ passing through the loop $C$ can only
have quantized values. We note that in our idealized considerations the loop
path $C$ is away from the boundaries.

We can prove $F(C)=1$ more explicitly by observing some simple properties of
$f({\bf a})$:

1. As $f({\bf a})$ is independent of ${\bf r}$, we can choose to explore the
properties of $f({\bf a})$ by considering displacements in regions where $%
{\bf B}=0$.

2. In any region of ${\bf B}=0$, consider the succession of two
infinitesimal displacements ${\bf a}$ and ${\bf b}$, as well as the
displacement ${\bf a}+{\bf b}$. Because ${\bf B}=0$ in the region enclosed
by the three vectors ${\bf a}$, ${\bf b}$ and ${\bf a}+{\bf b}$, it is clear
from (15) that
$$
f({\bf a}+{\bf b})=f({\bf a})f({\bf b}).\eqno(24)
$$

(24) offers another more explicit proof of (21). Recall that $F(C)$ is the
product of the f-factors along the loop path. Hence, for a closed loop, on
account of (24), we again obtain
$$
F(C)=f({\bf a})f({\bf b})f({\bf c})\cdots
$$
$$
=f({\bf a}+{\bf b}+{\bf c}+\cdots )
$$
$$
=f(\oint_Cd{\bf r})=f(0)=1.
$$
\\

\centerline{\bf Inclusion of Lattice Potential}

The lattice potential $V_{c}({\bf r})$, which is not invariant under
arbitrary displacements, has so far been omitted in our consideration. We
now consider the case when it is included.

Let the symmetry of the lattice be defined by three basic lattice vectors: $%
{\bf e}_1$, ${\bf e}_2$, ${\bf e}_3$. The lattice potential $V_c({\bf r})$
is then invariant under the space displacement
$$
{\bf r}_j\rightarrow {\bf r}_j-\triangle {\bf r,}\eqno(25)
$$
if$^9$
$$
\triangle {\bf r}=n_1{\bf e}_1+n_2{\bf e}_2+n_3{\bf e}%
_3,~~~~n_1,n_2,n_3=integers.\eqno(26)
$$
Under such displacements, the total Hamiltonian is invariant, except for an
induced gauge term in the vector potential, like that in (4). This is
exactly like the situation we had when the lattice potential was omitted,
except, of course, that the displacements have to be restricted to those in
(26). All previous arguments apply, leading to the conclusion, similar to
(19), that
$$
{\bf B}\cdot ({\bf e}_j\times {\bf e}_k)=n_{jk}\frac{ch}{2e}%
,~~~~n_{jk}=integers\eqno(27)
$$
which, in turn, implies that
$$
{\bf B}=\frac 1V\frac{ch}{2e}\sum_jn_j{\bf e}_j,~~~~n_j=integers\eqno(28)
$$
where
$$
V={\bf e}_1\cdot ({\bf e}_2\times {\bf e}_3).
$$
Namely, the magnetic field ${\bf B}$, if non-vanishing, inside the material
can only be along a discrete set of directions. Now, following Sewell$^3$,
we make the assumption that ${\bf B}$ is along the direction of the external
applied field. As the direction of the external field can be arbitrarily
chosen, the magnetic field ${\bf B}$ inside the material has to vanish, if
not to contradict (28), giving rise to the Meissner effect.

In the above proof we have made the assumption that the direction of ${\bf B}
$ inside the material lies along the direction of the external field. We
can, however, present another argument without having to make this
assumption. Again, consider the requirement imposed by (28). Typical lattice
dimension of superconducting materials is on the order of $1\AA $. The
smallest value of $B$, if not zero, has to be on the order of
$$
B\sim (1\AA )^{-2}\frac{ch}{2e}\sim 10^9G,\eqno(29)
$$
according to (28). This is many orders stronger than the typical critical
field strengths of less than $10^3G$, beyond which the superconducting state
no longer exists, for superconductors. Therefore, to be consistent with
(28), the only possible value of ${\bf B}$ is zero for a superconducting
state.

The same reasoning that lead to the quantization condition (23) applies, and
leads to
$$
\sum_C{\bf A}({\bf r})\cdot \triangle {\bf r}=n\frac{ch}{2e},~~~n=integers,%
\eqno(30)
$$
where loop path $C$ consists of a succession of displacements of the type
specified by (26). This quantization condition, when applied to
multiply-connected paths, implies the phenomenon of flux quantization. \\

\centerline{\bf Discussion}

We conclude with a few remarks.

1. The function $\Phi ({\bf r},{\bf r})$ corresponds to the Ginsburg-Landau
wave function$^{10}$. The successive displacements taken in arriving at
equation (16) clearly demonstrates the origin of the characteristics of the
superconducting state: Coherence of phases is responsible for the properties
of flux quantization and the Meissner effect. It is also made clear by the
proof given in this note that these twin properties of superconductivity
have a common origin, an understanding perhaps not sufficiently emphasized
in usual presentations.

2. The aspect in the hypothesis of ODLRO that has been made use of in the
proof of flux quantization and the Meissner effect is the factorization
property$^2$ of $\rho_{2}$ as in (13), which is responsible for the coherence
of
phases in $\Phi ({\bf r}_1,{\bf r}_2)$. The ODLRO hypothesis implicitly
assumes that in the sum in (12) there is only one term, having a large
eigenvalue on the order of $O(N)$, that survives in the ODLR limit.\\

\centerline{\bf Acknowledgement}

One of the authors (HTN) has benefited from discussions with A. S.
Goldhaber, J. K. Jain, V. Korepin, T. K. Ng and W. Weisberger. We would also
like to thank Prof. C. N. Yang, from whom they have learned over the years
the fundamentals of superconductivity. Gang Su is grateful to Prof. C. N.
Yang and ITP for the kind hospitality during his visit, and the Charles C.
K. Yeung fellowship through the CEEC of SUNY at Stony Brook for support. Two
of the authors (GS and BHZ) are supported in part by the National Natural
Science Foundation of China. \\

\centerline{\bf Footnotes}

\begin{description}
\item  $^1$O. Penrose, Phil. Mag. {\bf 42}, 1373(1951); O. Penrose and L.
Onsager, Phys. Rev. {\bf 104}, 576(1956).

\item  $^2$C. N. Yang, Rev. Mod. Phys. {\bf 34}, 694(1962).

\item  $^3$G. L. Sewell, J. Stat. Phys. {\bf 61}, 415(1990).

\item  $^4$ In proving the flux quantization, we shall need to consider the
case where there are regions inaccessible to the electrons, like a
tunnel. In these inaccessible regions, the magnetic field does
not necessarily vanish. In order to guarantee the single-valuedness of phase
$\chi _{{\bf a}}({\bf r}_j)$ in (7), we need to restrict to infinitesimal
{\bf a}, so that $\phi ({\bf r}_j-a)-\phi ({\bf r}_j)$ is single-valued,
although $\phi ({\bf r}_j-a)$ and $\phi ({\bf r}_j)$ individually are not.

\item  $^5$Here $\Phi ({\bf r}_1,{\bf r}_2)$ has been renormalized,  i.e., a
factor $
\sqrt{\alpha _0}$ has been absorbed into it.

\item  $^6$We thank W. Weisberger for pointing out the possible existence of
this displacement-dependent, but ${\bf r}$-independent phase factor.

\item  $^7$The consideration parallels Sewell's, in Ref. 3 , in his
derivation of the Meissner effect.

\item  $^8$The total flux includes the flux through the tunnel as well as
the flux of the penetrated magnetic field into the boundary layer
surrounding the tunnel.

\item  $^9$For reasons explained in footnote$^4$, we only consider $\Delta
{\bf r}$ which is infinitesimal on macroscopic scale. Thus, the integers $%
n_1,\,\,n_2\,\,$and $n_3$ can not be large.

\item  $^{10}$It is the factorization property that was assumed by L.P. Gor'kov
[JETP {\bf 7}, 505(1958)] to derive his effective equations for
superconductivity.
\end{description}

\end{document}